\documentclass[10pt,twocolumns,twoside,]{IEEEtran}
\usepackage{ntheorem}
\usepackage{amsmath}
\usepackage{amsfonts}
\newtheorem{theorem}{Theorem}
\newtheorem{corollary}{Corollary}
\newtheorem{lemma}{Lemma}
\newtheorem{definition}{Definition}
\newtheorem{assumption}{Assumption}
\newtheorem{remark}{Remark}
\usepackage{ntheorem}
\usepackage{amsmath}
\usepackage{amsfonts}
\usepackage{listings}
\usepackage{color}
\usepackage{array}
\usepackage{tabu}
\usepackage{mathrsfs,amssymb}
\usepackage{graphicx}
\graphicspath{ {image22/} }
\usepackage{float}
\usepackage{algorithm}
\usepackage{subcaption}

{}

\usepackage{algpseudocode}
\algdef{SE}[DOWHILE]{Do}{doWhile}{\algorithmicdo}[1]{\algorithmicwhile\ #1}%
\DeclareMathOperator*{\argmin}{arg\,min}

\DeclareMathOperator*{\card}{card}
\DeclareMathOperator*{\supp}{supp}

\begin{document}

	\title{Secure Estimation and Attack Isolation for Connected and Automated Driving in the Presence of Malicious Vehicles}
	
	%at, but, by, for, in, nor, of, on, or, the, to and up, which are usually
	%\title{An Unknown Input Multi-Observer Approach for Estimation, Attack Isolation, and Control of LTI Systems under Adversarial Attacks}
	
	\author{Tianci~Yang,
	\and~Chen~Lv,
	\emph{Senior Member, IEEE}% <-this % stops a space
	\thanks{This work was supported by the SUG-NAP Grant (No. M4082268.050) of Nanyang Technological University, Singapore.}% <-this % stops a space
	\thanks{The authors are with the School of Mechanical and Aerospace Engineering, Nanyang Technological University, Singapore.
		\{tianci.yang, lyuchen\}@ntu.edu.sg}%%
}

	\markboth{Journal of \LaTeX\ Class Files,~Vol.~14, No.~8, April~2020}%
	{Shell \MakeLowercase{\textit{et al.}}: Bare Demo of IEEEtran.cls for IEEE Journals}



	\maketitle
	\begin{abstract}
		Connected and Automated Vehicles (CAVs) rely on the correctness of position and other vehicle kinematics information to fulfill various driving tasks such as vehicle following, lane change, and collision avoidance. However, a malicious vehicle may send false sensor information to the other vehicles intentionally or unintentionally, which may cause traffic inconvenience or loss of human lives. Here, we take the advantage of cloud-computing and increase the resilience of CAVs to malicious vehicles by assuming each vehicle shares its local sensor information with other vehicles to create information redundancy on the cloud side. We exploit this redundancy and propose a sensor fusion algorithm for the cloud, capable of providing a robust state estimation of all vehicles in the cloud under the condition that the number of malicious information is sufficiently small. Using the proposed estimator, we provide an algorithm for isolating malicious vehicles. We use numerical examples to illustrate the effectiveness of our methods.
	\end{abstract}
\begin{IEEEkeywords}
	Secure estimation, attack isolation, connected and automated vehicles, sensor fusion, malicious vehicles.
\end{IEEEkeywords}
	\section{Introduction}
	In the recent years, the increasing number and length of the traffic congestion caused by limited highway capacity has become a main concern. With the development of wireless communication and sensing technologies, connected and automated vehicles (CAVs) are expected to enhance the safety, efficiency and throughput of our intelligent transportation systems via sharing its local sensor measurements with the other surrounding vehicles through vehicular networks. However, the sensors of modern vehicles are usually designed without enough security concerns and hence remain vulnerable to cyberattacks; moreover, because of a lack of centralized administration, vehicular networks are also providing chances for attackers to disrupt the normal operation of the transportation systems, see, e.g., \cite{El-rewini2020}-\nocite{yang2019deqos}\nocite{Laurendeau2006}\nocite{car}\nocite{Creenberg}\nocite{Amoozadeh2015}\nocite{amoozadeh2015security}\nocite{ju2020distributed} \nocite{lu2014connected}\nocite{sanders2020localizing}\nocite{parkinson2017cyber}\nocite{litman2017autonomous}\nocite{anderson2014autonomous}\cite{wu2020towards}, and references therein. In \cite{Creenberg}, researchers showed that they were able to stop the engine of the vehicle remotely while it was driving down a busy highway. The vulnerabilities of millimeter-wave radars, ultrasonic sensors and forward-looking cameras to attacks is investigated in \cite{yan2016can}. The authors of \cite{petit2015remote} show that various types of attacks can be performed using commonly affordable commodity hardware. A simpler attacks on vehicular networks might be a malicious vehicle transmitting fraudulent data about road conditions or vehicle positions for its own profits, which might cause traffic jams or catastrophic consequences. Cyberattacks on a connected vehicle are severely threatening not only the normal operation of our transportation systems, but also the lives of the drivers, the passengers and the pedestrians. This has shown that great efforts needs to be made to provide strategic mechanism for identifying and mitigating cyberattacks on CAVs.
	
	In the literature of CAV security, most of the existing results give their solutions using cryptography-based methods \cite{Chen2019}-\nocite{contreras2017internet}\nocite{mundhenk2017security}\nocite{8939382}\nocite{hartkopp2012message}\nocite{ raya2007securing}\cite{hasrouny2017vanet}, which mainly aim to prevent in-vehicle or inter-vehicle networks from being attacked using different authentication and authorization protocols. There has not been much research on providing methods for mitigating the performance degradation resulting from cyberattacks on CAVs. In \cite{yan2008providing}, the authors provide an algorithm for malicious vehicles detection mainly by comparing its local on-board sensor measurements with the information received from other vehicles. An algorithm for detecting malicious vehicles was proposed in \cite{daeinabi2013detection}, where it is assumed that the vehicle cluster has leaders providing reliable information. The authors of \cite{barnwal2012heartbeat} propose a misbehavior detection scheme for identifying the source of false information, assuming that each vehicle is able to predict other vehicles' behavior and a misbehavior is detected when the prediction and the received information from other vehicles mismatch. Here, we propose a sensor fusion algorithm for connected and automated driving in the presence of malicious vehicles, capable of providing a robust state estimation for each vehicle and efficiently pinpointing malicious vehicles. 

	Our sensor fusion algorithm is inspired by \cite{Chong2015}, where a state estimator is constructed for general linear time-invariant (LTI) systems under sensor attacks using a bank of observers. Placing redundant sensors/actuators in systems and then exploiting sensor/actuator redundancy is a commonly adopted technique in the research area of secure estimation and control for cyber-physical systems \cite{Chong2015}-\nocite{Fawzi2014}\nocite{Shoukry2017}\nocite{Kim2016a}\nocite{Yang2018a}\cite{ yang2020multi}. In this manuscript, we create sensor redundancy by assuming each CAV collaborates with each other by sharing its local sensor measurements via the cloud. We consider each vehicle is measuring its own state and also the relative state between itself and its surrounding vehicles using its on-board sensors, and then upload the information to a cloud so that redundancy of the state of each vehicle is created on the cloud side. However, some malicious vehicles are uploading corrupted data intentionally or unintentionally and we do not know how many vehicles or which of them are malicious. Note that the information of the state of each vehicle uploaded by different vehicles will also be different in general even in the absence of malicious vehicles because of the sensor noise and the imperfections induced by the network effects (delay, quantization, packet dropouts .etc). This poses a challenge to the cloud for distinguishing between the healthy data uploaded by attack-free vehicles and the corrupted data uploaded by malicious vehicles. Our sensor fusion algorithm overcomes this challenge, capable of providing robust estimates of the true state of each vehicle in the cloud from all the collected information. The algorithm guarantees an attack-free state estimation provided that less than half of the collected information is under attack. After the cloud obtain attack-free state estimate of each vehicle, if an upper bound on the disturbance is known, an isolation algorithm can be designed for pinpointing the malicious vehicles in the cloud. We determine whether a vehicle is malicious or not by computing the differences between the estimates we have and the corresponding information uploaded by this vehicle: if a vehicle is attack-free, all the information it uploads will be consistent with the estimates obtained by the cloud, and hence the differences will be smaller than a pre-determined threshold; otherwise, the difference might cross the threshold and the vehicle is isolated as a malicious one. Thanks to the power of cloud computing \cite{ibanez2015integration}-\nocite{whaiduzzaman2014survey}\cite{hussain2012rethinking}, the computation burden of secure state estimation can be moved from CAVs to the cloud. To increase the robustness of our algorithm to Sybil attacks \cite{douceur2002sybil}, where multiple malicious identities in the cloud can be created by a single attacker, the methodologies provided in the literature \cite{levine2006survey}-\nocite{guette2007sybil}\cite{chen2009robust} can be used for detecting Sybil attacks before running our algorithm. 
	
	There are a few results in this direction already. The problem of attack detection and state estimation for connected vehicles under sensor attack is addressed in \cite{ju2020deception} using Unbiased Finite Impulse Response (UFIR) filters. However, their results are only applicable to the case when sensor attack signals are piece-wise constant. In \cite{golle2004detecting}, the authors propose to correct the malicious data in vehicular networks by finding the ``best explanation'' for the data received. The authors of \cite{chattopadhyay2018secure} provide an adaptive filtering approach for state estimation in the presence of malicious data.
	 Unfortunately, the performance of the estimation scheme highly depends on the choice of several tuning parameters, which leads to incorrect estimation if chosen inappropriately, and the method for choosing these parameters correctly are not explicitly given in \cite{chattopadhyay2018secure}. More importantly, there is no guaranteed estimation performance for the methods provided in \cite{golle2004detecting}\cite{chattopadhyay2018secure}.
	
	The main contributions of this manuscript are the following: 1) we give sufficient and necessary conditions under which the state of each vehicle can be reconstructed from the corrupted measurements without ambiguity; 2) we propose a sensor fusion algorithm for providing a robust state estimation for each vehicle in the cloud in spite of malicious vehicles -- with guaranteed estimation performance -- without any prior knowledge of system disturbances is known except that they are bounded -- without any restrictions on the injected attack signals; 3) we provide an effective algorithm for isolating the malicious vehicles.
	
	The paper is organized as follows. Notations are given in Section \ref{preliminary}. In Section \ref{formulation}, we formulate the problem we consider. In Section \ref{fusion}, a secure sensor fusion algorithm is proposed in the presence of malicious vehicles. An algorithm for isolating malicious vehicles is presented in Section \ref{isolation}. In Section \ref{simulation}, numerical examples are used to demonstrate the performance of our algorithms. Finally, concluding remarks are given in Section \ref{conclusion}.
\section{Preliminary}\label{preliminary}
	The following standard notations are used in this manuscript.
	We denote the set of real numbers by $\mathbb{R}$, the set of natural numbers by $\mathbb{N}$, and $\mathbb{R}^{n\times m}$ the set of $n\times m$ matrices for any $m,n \in \mathbb{N}$. For any vector $v\in\mathbb{R}^{n}$,  we denote {$v^{J}$} the stacking of all $v_{i}$, $i\in J$, $J\subset \left\lbrace 1,\hdots,n\right\rbrace$, $|v|=\sqrt{v^{\top} v}$, and $\supp(v)=\left\lbrace i\in\left\lbrace 1,\hdots,n\right\rbrace |v_{i}\neq0\right\rbrace $. For a sequence $\left\lbrace v(k)\right\rbrace _{k=0}^{\infty}$,  $||v||_{\infty} := \sup_{k\geq 0}|v(k)|$, $v(k) \in \mathbb{R}^{n}$. We say that signal $\left\lbrace v(k)\right\rbrace$ belongs to $l_{\infty}$, $\left\lbrace v(k)\right\rbrace \in l_{\infty}$, if $||v||_{\infty}<\infty$. We denote the cardinality of a set $S$ as $\card(S)$. We denote a variable $m$ uniformly distributed in the interval $(z_{1},z_{2})$ as $m\sim\mathcal{U}(z_{1},z_{2})$ and normally distributed with mean $\mu$ and variance $\sigma^2$ as $m\sim \mathcal{N}(\mu,\sigma^2)$.
	\section{Problem Formulation}\label{formulation}
\begin{figure}[t]\centering
		\includegraphics[width=0.5\textwidth]{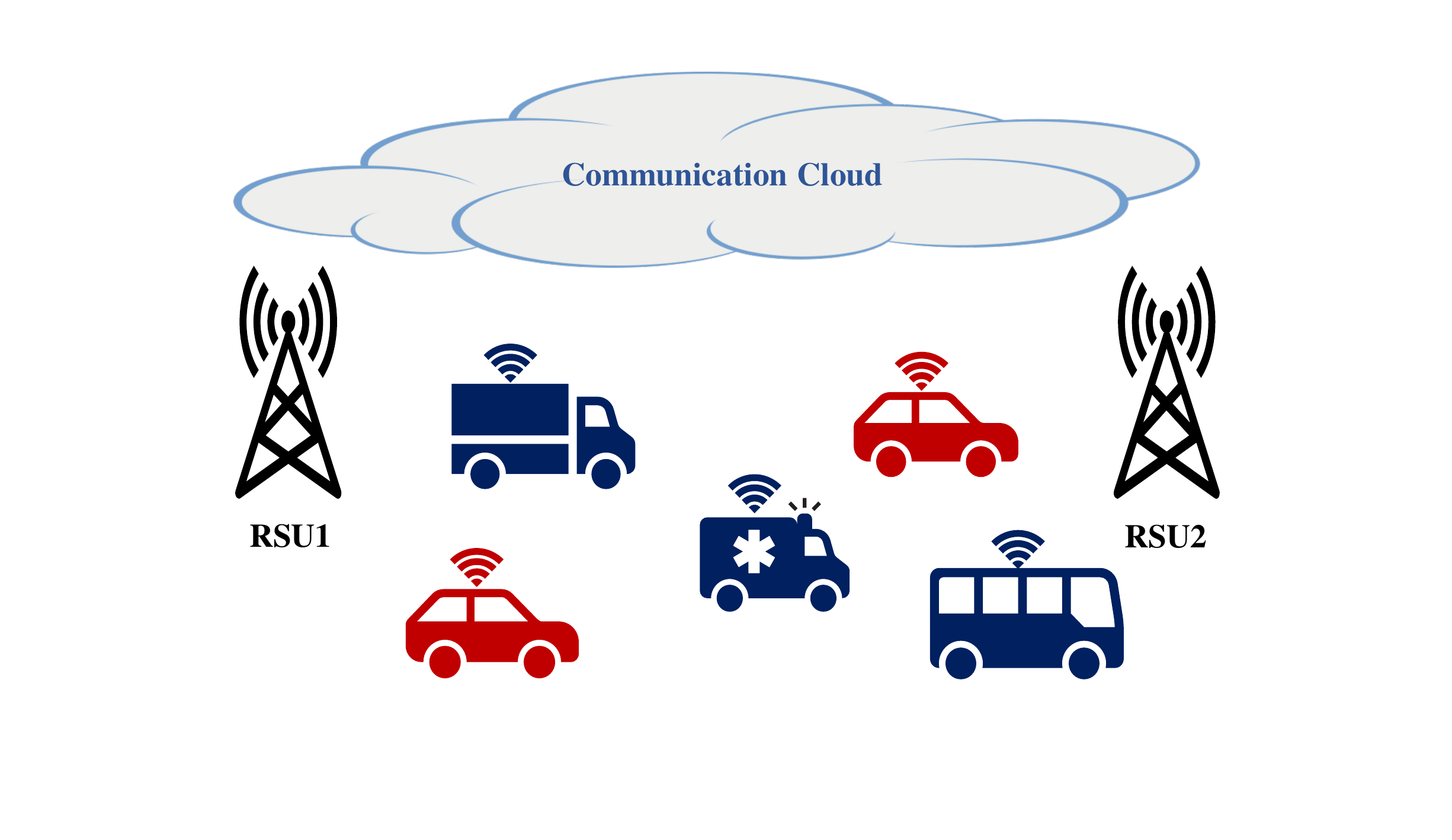}
		\caption{Attack-free vehicles (black) upload its local sensor measurements to the cloud; malicious vehicles (red) upload manipulated sensor measurements to the cloud.}
		\centering
		\label{fig:0}
	\end{figure}

	We consider a group of connected and automated vehicles (CAVs) as depicted in Figure \ref{fig:0}, each vehicle in the group uploads its own state and also the relative state between itself and its surrounding vehicles to the cloud at every time step. However, some of the information stored in the cloud might be corrupted since some vehicles in the group might be malicious, e.g., a malicious vehicle may be driven by an attacker such that manipulated sensor information is uploaded intentionally on purpose of his own selfish profits; a malicious vehicle might also be driven by an innocent driver without noticing that its on-board sensors are faulty or compromised by a remote attacker \cite{petit2015remote}. 
	Denote the set of all vehicles in the cloud as $S(k)$ for $k\geq 0$. Suppose the state of vehicle $i\in S(k)$ at time $k\geq 0$ is denoted as $x_{i}(k)\in\mathbb{R}$, which can represent, for instance, the velocity or position of vehicle $i$. For $k\geq 0$, the cloud collects $x_{i}(k)$ which is uploaded by vehicle $i$ and the received information is denoted as $\tilde{x}_{i}(k)$ with 
	\begin{equation}
		\tilde{x}_{i}(k)=x_{i}(k)+m_{i}(k)+a_{i}(k),
	\end{equation}
	where $m_{i}$ is the disturbance caused by network effects (quantization, delay, packet dropouts etc.) and sensor noise, $a_{i}$ is the malicious signal injected to vehicle $i$ sensors, i.e., $a_{i}(k)\neq 0$ for some $k\geq 0$ and can take any arbitrary value if vehicle $i$ is malicious; otherwise, $a_{i}(k)=0$ for all $k\geq 0$. Let $\mathcal{N}_{i}(k)$ represent the set of surrounding vehicles of vehicle $i$ at time $k$. The cloud also collects the measurements of $x_{i}(k)$ uploaded by vehicle $j$ for all $j\in\mathcal{N}_{i}(k)$ to create information redundancy:
	\begin{equation}\label{xij}
		\tilde{x}_{i|j}(k)=x_{j}(k)+r_{i|j}(k)+m_{j}(k)+a_{j}(k),
	\end{equation}
	where $\tilde{x}_{i|j}\in\mathbb{R}$ represents the measurement of $x_{i}$ given by vehicle $j$, $x_{j}\in\mathbb{R}$ is the state of vehicle $j$, $r_{i|j}$ is the relative state between vehicles $j$ and $i$, $m_{j}$ is the disturbance caused by network effects and sensor noise, $a_{j}$ is the malicious signals, i.e., $a_{j}(k)\neq 0$ for some $k\geq 0$ and can take any arbitrary value if vehicle $j$ is malicious; otherwise, $a_{j}(k)=0$ for all $k\geq 0$.
	
	\begin{assumption}\label{as1}
		For all $i\in S(k)$ and $k\geq 0$, vehicle $i$ has at least two surrounding vehicles that measure $x_{i}(k)$ at time $k$, i.e., $\card(\mathcal{N}_{i}(k))\geq 2$.
	\end{assumption}
	Under Assumption \ref{as1}, $N_{i}(k)$ copies of $x_{i}(k)$ are collected on the cloud side with $N_{i}(k)=1+\card(\mathcal{N}_{i}(k)$ at time $k$ with $N_{i}(k)\geq 3$, which is denoted as follows
	\begin{equation}
		\begin{split}
			X_{i}(k)=&\begin{bmatrix}
				x_{i}(k)\\
				x_{i}^{\mathcal{N}_{i}(k)}(k)
			\end{bmatrix}+m(k)+a(k),\\
			=&\begin{bmatrix}
				x_{i}(k)\\
				\vdots\\
				x_{i}(k)\\
				\vdots\\
				x_{i}(k)
			\end{bmatrix}+m(k)+a(k)=\begin{bmatrix}
				X_{i1}(k)\\
				\vdots\\
				X_{ij}(k)\\
				\vdots\\
				X_{iN_{i}(k)}(k)
			\end{bmatrix}.
		\end{split}
	\end{equation}
	where $X_{i}\in\mathbb{R}^{N_{i}(k)}$, $x_{i}^{\mathcal{N}_{i}(k)}\in\mathbb{R}^{\card(\mathcal{N}_{i}(k)}$denotes the stacking of $x_{j}+r_{i|j}$, for all $j\in\mathcal{N}_{i}(k)$, $m=\begin{bmatrix}
		m_{i}&m^{\mathcal{N}_{i}(k)}
	\end{bmatrix}^{\top}$, $a=\begin{bmatrix}
		a_{i}&a^{\mathcal{N}_{i}(k)}
	\end{bmatrix}^{\top}$ with $m^{\mathcal{N}_{i}(k)}$ denotes the stacking of $m_{j}$ for all $j\in \mathcal{N}_{i}(k)$, and $a^{\mathcal{N}_{i}(k)}$ denotes the stacking of $a_{j}$ for all $j\in {\mathcal{N}_{i}(k)}$. For $j\in\left\lbrace 1,\ldots,N_{i}(k)\right\rbrace $, if the $j$-th vehicle is attack-free, then $a_{j}(k)=0$ for all $k\geq 0$; otherwise $a_{j}(k)\neq 0$ for some $k\geq 0$ and can take any arbitrary value. For all $i\in S(k)$ and $k\geq 0$, define $\mathcal{N}_{i+}(k)\triangleq\mathcal{N}_{i}(k)\cup \left\lbrace i\right\rbrace $ and denote the set of malicious information as $W_{i}(k)\subset\left\lbrace 1,\ldots, N_{i}(k)\right\rbrace  $, i.e., 
	\begin{equation}
		\supp(a(k))\subseteq W_{i}(k),\hspace{2mm}k\geq 0.
	\end{equation}
	\begin{assumption}\label{as2}
		The sensor disturbance caused by network effects and sensor noise is bounded, i.e., $\left\lbrace m_{i}(k)\right\rbrace \in l_{\infty}$. And the values of $||m_{i}||_{\infty}$ is unknown for any $i\in\left\lbrace 1,\ldots,N_{i}(k)\right\rbrace $ and $k\geq 0$.
	\end{assumption}

\begin{figure}[t]\centering
	\includegraphics[width=0.5\textwidth]{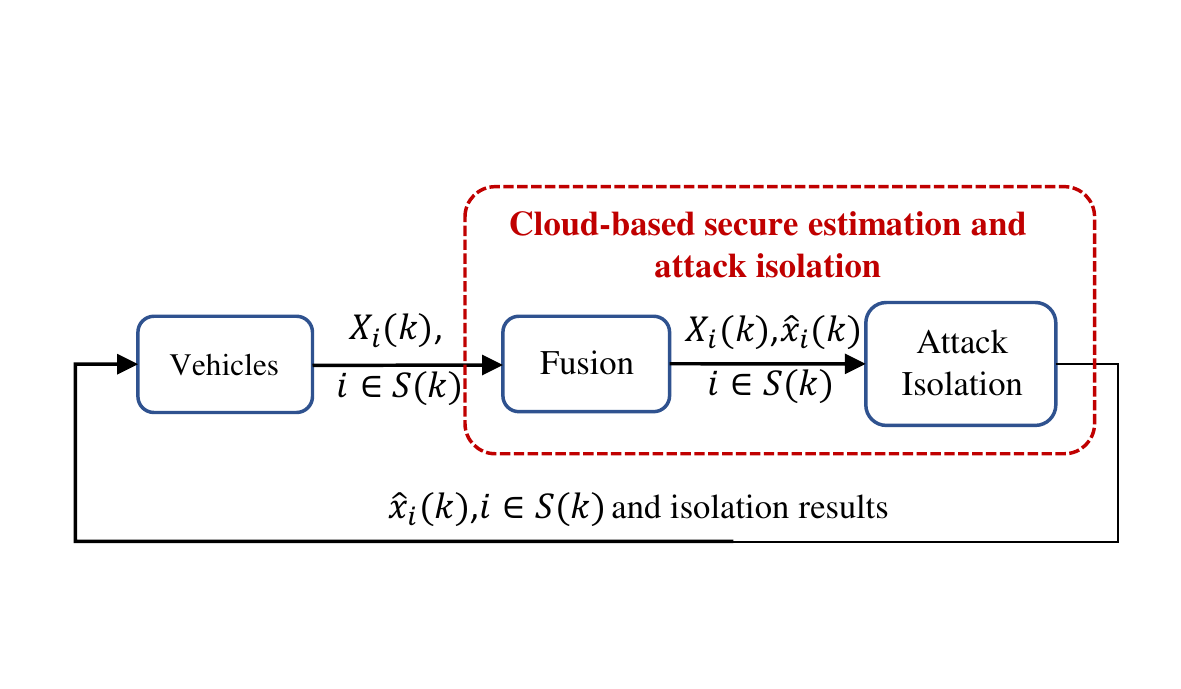}
	\caption{Schematic representation of the system.}
	\centering
	\label{fig:block}
\end{figure}

\textbf{Problems:} We first aim to investigate the sufficient and necessary conditions under which $x_{i}(k)$ can be reconstructed from the corrupted measurements $X_{i}(k)$ without ambiguity for $i\in S(k)$ and $k\geq 0$, and design a sensor fusion-based estimation scheme for the cloud that is capable of reconstructing $x_{i}(k)$ from $X_{i}(k)$ for all $i\in S(k)$ and $k\geq 0$. Then we focus on the design of an algorithm for isolating malicious vehicles using the proposed estimator while assuming that upper bounds on the system disturbances are available. A schematic representation of our system is dipicted in Figure \ref{fig:block}.

		\section{Secure Sensor Fusion}\label{fusion}
		In this section, for each vehicle $i\in S(k)$, we first provide the sufficient and necessary condition for $x_{i}(k)$ to be reconstructible from the malicious measurements $X_{i}(k)$, then develop a sensor fusion algorithm for providing a state estimation of each vehicle independent of the attack signals.
			\begin{definition}[Reconstructability]
			$x_{i}(k)$ is reconstructible from $X_{i}(k)$ under $q_{i}(k)$ attacks at time $k\geq 0$, if for all $x_{i}(k)$, $\bar{x}_{i}(k)$, sets $W_{1}(k)$, $W_{2}(k)\subset\left\lbrace 1,\ldots, N_{i}(k)\right\rbrace $ with $\card(W_{1}(k))$, $\card(W_{2}(k))\leq q_{i}(k)$, and $\supp(a_{1}(k))\subseteq W_{1}(k)$, $\supp(a_{2}(k))\subseteq W_{2}(k) $ and some $k\geq 0$, we have
			\begin{equation}
				X_{i}(k)=\bar{X}_{i}(k)\Longrightarrow x_{i}(k)=\bar{x}_{i}(k).
			\end{equation}
		\end{definition}
		
		The meaning of this definition is that when $x_{i}(k)$ is reconstructible from $X_{i}(k)$ under $q_{i}(k)$ attacks at time $k$, then there exists a unique $x_{i}(k)$ that can explain the data collected by the cloud when the number of corrupted information is less than $q_{i}(k)$ at time $k$, and hence $x_{i}(k)$ can be reconstructed from $X_{i}(k)$ unambiguously at time $k$.
		\begin{theorem}\label{b}
			$x_{i}(k)$ is reconstructible from $X_{i}(k)$ under $q_{i}(k)$ attacks for all $k\geq 0$ if and only if $q_{i}(k)<\frac{N_{i}(k)}{2}$ for all $k\geq 0$.
		\end{theorem}
		\textit{Proof:}
		
		\textbf{1) if:}  Suppose that $q_{i}(k)<\frac{N_{i}(k)}{2}$ for all $k\geq 0$.  For all $a(k)$ and $\bar{a}(k)$ with $\card(\supp(a(k))\leq q_{i}(k)$ and $\card(\supp(\bar{a}(k)))\leq q_{i}(k)$, the nonzero elements of $\bar{a}(k)-a(k)$ is no more than $2q_{i}(k)$, which is strictly smaller than $N_{i}(k)$. Therefore, there do not exist $a(k)\neq \bar{a}(k)$ such that
		\begin{equation}\label{e}
			\bar{a}(k)-a(k)=\left[ \begin{matrix}
				x_{i}(k)-\bar{x}_{i}(k)\\
				x_{i}(k)-\bar{x}_{i}(k)\\
				\vdots\\
				x_{i}(k)-\bar{x}_{i}(k)
			\end{matrix}\right],
		\end{equation}
		with $x_{i}(k)\neq\bar{x}_{i}(k)$ for any $k\geq 0$, since the number of nonzero elements on the right side of \eqref{e} is equal to $N_{i}(k)$. Therefore, there do not exist $x_{i}(k)\neq\bar{x}_{i}(k)$ such that
		\begin{equation}\label{eq1}
			\left[ \begin{matrix}
				x_{i}(k)+m_{1}(k)\\
				x_{i}(k)+m_{2}(k)\\
				\vdots\\
				x_{i}(k)+m_{N_{i}(k)}(k)
			\end{matrix}\right] +a(k)=\left[ \begin{matrix}
				\bar{x}_{i}(k)+m_{1}(k)\\
				\bar{x}_{i}(k)+m_{2}(k)\\
				\vdots\\
				\bar{x}_{i}(k)+m_{N_{i}(k)}(k)
			\end{matrix}\right] +\bar{a}(k),
		\end{equation}
		for any $k\geq 0$, which indicates $x_{i}(k)$ is reconstructible from $X_{i}(k)$ for all $k\geq 0$.
		
		\textbf{2) only if:}
		We prove by contrapositive. Suppose $q_{i}(k_{1})\geq\frac{N_{i}(k_{1})}{2}$ for some $k_{1}\geq 0$, then for all $W(k_{1})\subset\left\lbrace 1, \ldots, N_{i}(k_{1})\right\rbrace $, with $\card(W(k_{1}))=q_{i}(k_{1})$, there must exist another set $\bar{W}(k_{1})\subset\left\lbrace 1, \ldots, N_{i}(k_{1})\right\rbrace $ with $\card(\bar{W}(k_{1}))=q_{i}(k_{1})$ such that $W(k_{1})\cup \bar{W}(k_{1})=\left\lbrace 1, \ldots, N_{i}(k_{1})\right\rbrace $ since $2q_{i}(k_{1})\geq N_{i}(k_{1})$. Let $I(k_{1})=W(k_{1})\cap \bar{W}(k_{1})$. For $x_{i}(k_{1})\neq \bar{x}_{i}(k_{1})$, we choose 
		\begin{equation}
			\begin{split}
				&\bar{a}_{j}(k_{1})=x_{i}(k_{1})-\bar{x}_{i}(k_{1}), a_{j}(k_{1})=0, j\in \left\lbrace \bar{W}(k_{1})\setminus I(k_{1})\right\rbrace \\
				&\bar{a}_{j}(k_{1})=0, a_{j}(k_{1})=\bar{x}_{i}(k_{1})-x_{i}(k_{1}), j\in \left\lbrace W(k_{1})\setminus I(k_{1})\right\rbrace \\
				&\bar{a}_{j}(k_{1})=x_{i}(k_{1}),a_{j}(k_{1})=\bar{x}_{i}(k_{1}), \hspace{6mm}j\in I(k_{1}).
			\end{split}
		\end{equation}
		\vspace{3mm}
		Then we have \eqref{e} holds at time $k_{1}$ with $x_{i}(k_{1})\neq \bar{x}_{i}(k_{1})$, which indicates
		\begin{equation}
			X_{i}(k_{1})=\bar{X}_{i}(k_{1}),
		\end{equation}
		with $x_{i}(k_{1})\neq \bar{x}_{i}(k_{1})$. This contradicts the statement that $x_{i}(k)$ is reconstructible from $X_{i}(k)$ under $q_{i}(k)$ attacks for all $k\geq 0$.\hfill$\blacksquare$ 
		\begin{assumption}\label{as3}
			For all $i\in S(k)$, in the set $\mathcal{N}_{i+}(k)$, the number of malicious information does not exceed $\frac{N_{i}(k)}{2}$ for all $k\geq 0$, i.e.,
			\begin{equation}
				\card(W_{i}(k))\leq q_{i}(k)<\frac{N_{i}(k)}{2}, k\geq 0.
			\end{equation}
		\end{assumption}
		\begin{remark}
			Under Assumption \ref{as3}, all vehicles in $\mathcal{N}_{i+}(k)$ can be malicious at some time $k\geq 0$, but at each time step $k$, the number of vehicles with nonzero malicious signals in $\mathcal{N}_{i+}(k)$ is no larger than $q_{i}(k)$. This limitation may result from the fact that the on-board sensors of each vehicle might be faulty at times, but it is more likely that only a small fraction of vehicles can have faulty sensors at the same time step; or it may result from limited capability of the attacker, or the attack strategy chosen by the adversary.  
		\end{remark}
		\begin{corollary}\label{c2}
			Under Assumption \ref{as3}, among all $N_{i}(k)$ measurements of $x_{i}(k)$, at least $N_{i}(k)-q_{i}(k)$ of them are attack-free; among every set of $N_{i}(k)-q_{i}(k)$ measurements of $x_{i}(k)$, at least $N_{i}(k)-2q_{i}(k)$ of them are attack-free.
		\end{corollary}
	For every subset $J\subset\left\lbrace 1, \ldots, N_{i}(k)\right\rbrace $ measurements and $k\geq 0$, we define $\hat{x}_{J}(k)$ as the average value of all the measurements given by subset $J$ at time $k$, as follows:
	\begin{equation}
		\hat{x}_{J}(k)\triangleq  \frac{\sum_{j\in J}X_{ij}(k)}{\card(J)}.
	\end{equation}
	Define
	\begin{equation}
		||m||_{\infty}\triangleq\max_{i\in\left\lbrace 1,\cdots,N_{i}(k)\right\rbrace,k\geq 0}\left\lbrace ||m_{i}||_{\infty}\right\rbrace.
	\end{equation} 
	\vspace{3mm}
	\begin{remark}
		We assume that $||m_{i}||_{\infty}$ and $||m||_{\infty}$ are unknown.
	\end{remark}
	\begin{lemma}
		If $a^{J}(k)=0$, then
		\begin{equation}
			|\hat{x}_{J}(k)-x_{i}(k)|\leq ||m||_{\infty}.
		\end{equation}
		for all $k\geq 0$.
	\end{lemma}
	\textit{Proof:}
	Since
	\begin{equation}
		\begin{split}
			&\left|\sum_{i}^{n}m_{i}\right|=\sqrt{(m_{1}+m_{2}+\cdots+m_{n})^{2}}\\
			&=\sqrt{m_{1}^{2}+m_{2}^{2}+\cdots+m_{n}^{2}+2(m_{1}m_{2}+\cdots+m_{n-1}m_{n})}\\
			&\leq\sqrt{m_{1}^{2}+m_{2}^{2}+\cdots+m_{n}^{2}+(m_{1}^{2}+m_{2}^{2}+\cdots+m_{n-1}^{2}+m_{n}^{2})}\\
			&=\sqrt{\left( 1+\left( \begin{matrix}
					1\\
					n-1
				\end{matrix}\right) \right) \left( m_{1}^{2}+m_{2}^{2}+\cdots+m_{n}^{2}\right\rbrace }\\
			&\leq\sqrt{n^{2}||m||_{\infty}^{2}}\\
			&=n||m||_{\infty},
		\end{split}
	\end{equation}
	we have
	\begin{equation}
		\begin{split}
			|\hat{x}_{J}(k)-x_{i}(k)|=&\left| \frac{\sum_{j\in J}X_{ij}(k)}{\card(J)}-x_{i}(k)\right|\\
			=&\frac{1}{\card(J)}\left|\sum_{j\in J}m_{j}(k)\right|\\
			\leq&\frac{1}{\card(J)}\card(J)||m||_{\infty}\\
			=&||m||_{\infty}.
		\end{split}
	\end{equation}
	\hfill$\blacksquare$
	
	Under Assumption \ref{as3}, there exists at least one subset $\bar{I}(k)\subset \left\lbrace 1, \ldots, N_{i}(k)\right\rbrace $ with $\card(\bar{I}(k))=N_{i}(k)-q_{i}(k)$ such that $a^{\bar{I}(k)}(k)=0$ for $k\geq 0$. Then, in general, the difference between $\hat{x}_{\bar{I}(k)}(k)$ and any $X_{ij}(k)$, $j\in\bar{I}(k)$ will be smaller than the other subsets $J\subset\left\lbrace 1, \ldots, N_{i}(k)\right\rbrace $ with $\card(J)=N_{i}(k)-q_{i}(k)$ and $a^{J}(k)\neq 0$. This motivates the following fusion algorithm.

	For every subset $J\subset\left\lbrace 1,\ldots,N_{i}(k)\right\rbrace $ of measurements with $\card(J)=N_{i}(k)-q_{i}(k)$, define $\pi_{J}(k)$ as the largest difference between $\hat{x}_{J}(k)$ and $X_{ij}(k)$ for all $j\in J$, i.e.,
	\begin{equation}\label{es1}
		\pi_{J}(k)=\underset{j\in J}{\max}\left| \hat{x}_{J}(k)-X_{ij}(k)\right| 
	\end{equation}
	for all $k\geq 0$, and the sequence $\sigma(k)$ is given as,
	\begin{equation}\label{es2}
		\begin{split}
			\sigma(k)=\underset{J\subset\left\lbrace 1,\ldots, N_{i}(k)\right\rbrace: \card(J)=N_{i}(k)-q_{i}(k)}{\argmin}\pi_{J}(k).
		\end{split}
	\end{equation}
	Then, as proved below, the estimate indexed by $\sigma(k)$:
	\begin{equation}\label{es3}
		\hat{x}_{i}(k)=\hat{x}_{\sigma(k)}(k),
	\end{equation}
	is an attack-free estimate of $x_{i}(k)$. 
	\begin{theorem}
		Consider the fusion algorithm \eqref{es1}-\eqref{es3}. Define the estimation error $e_{i}(k):=\hat{x}_{\sigma(k)}(k)-x_{i}(k)$ for all $i\in S(k)$ and $k\geq 0$, and let Assumptions \ref{as1}-\ref{as3} be satisfied; then,
		\begin{equation}
			|e_{i}(k)|\leq 3||m||_{\infty}\label{sa},
		\end{equation}
		for all $i\in S(k)$ and $k\geq 0$.
	\end{theorem}
	\textit{Proof:}
	Under Assumption \ref{as3}, there exists at least one subset $\bar{I}(k)\subset\left\lbrace 1, \ldots, N_{i}(k)\right\rbrace $ with $\card(\bar{I}(k))=N_{i}(k)-q_{i}(k)$ such that $a^{\bar{I}(k)}(k)=0$ for $k\geq 0$. Then 
	\begin{equation}
		|\hat{x}_{\bar{I}(k)}(k)-x_{i}(k)|\leq ||m||_{\infty}.
	\end{equation}
	Moreover, for all $j\in\bar{I}(k)$, $a_{j}(k)=0$ and we have
	\begin{equation}
		|X_{ij}(k)-x_{i}(k)|=|m_{j}(k)|.
	\end{equation}
	Then we have
	\begin{equation}
		\begin{split}
			\pi_{\bar{I}(k)}(k)=&\underset{j\in \bar{I}(k)}{\max}\left| \hat{x}_{\bar{I}(k)}(k)-X_{ij}(k)\right| \\
			=&\underset{j\in \bar{I}(k)}{\max}\left|\hat{x}_{\bar{I}(k)}(k)-x_{i}(k)+x_{i}(k)-X_{ij}(k)\right| \\
			=&\left| \hat{x}_{\bar{I}(k)}(k)-x_{i}(k)\right| +\underset{j\in \bar{I}(k)}{\max}\left| x_{i}(k)-X_{ij}(k)\right| \\
			\leq&||m||_{\infty}+\underset{j\in \bar{I}(k)}{\max}|m_{j}(k)|.
		\end{split}
	\end{equation}
	%since
	%\begin{equation}
	%\begin{split}
	%\left|\sum_{i=1}^{3}\nu_{i}\right|=&\sqrt{(\nu_{1}+\nu_{2}+\nu_{3})^{2}}\\
	%=&\sqrt{\nu_{1}^{2}+\nu_{2}^{2}+\nu_{3}^{2}+2\nu_{1}\nu_{2}+2\nu_{1}\nu_{3}+2\nu_{2}\nu_{3}}\\
	%\leq&\sqrt{\nu_{1}^{2}+\nu_{2}^{2}+\nu_{3}^{2}+\nu_{1}^{2}+\nu_{2}^{2}+\nu_{1}^{2}+\nu_{3}^{2}+\nu_{2}^{2}+\nu_{3}^{2}}\\
	%=&\sqrt{3}\sqrt{\nu_{1}^{2}+\nu_{2}^{2}+\nu_{3}^{2}}\\
	%\leq&\sqrt{3}\sqrt{3||\nu||_{\infty}^{2}}
	%=3||\nu||_{\infty}
	%\end{split}
	%\end{equation}
	%and we also have
	%\begin{equation}
	%\left| \frac{\sum_{i\in \bar{S}}U_{i}(k)}{N-2q}-u_{i-1}(k)\right| =\frac{\sqrt{N-2q}}{N-2q}|\nu^{\bar{S}}(k)|
	%\end{equation} 
	From Corollary \ref{c2}, among every set of $N_{i}(k)-q_{i}(k)$ vehicles, at least one of the vehicles is attack free since $N_{i}(k)-2q_{i}(k)\geq 1$. Therefore, there exists at least one $\bar{j}(k)\in\sigma(k)$ such that $a_{\bar{j}(k)}(k)=0$ for $k\geq 0$ and
	\begin{equation}
		\begin{split}
			|X_{i\bar{j}(k)}(k)-x_{i}(k)|=|m_{\bar{j}(k)}(k)|.
		\end{split}
	\end{equation}
	From \eqref{es2}, we have $\pi_{\sigma(k)}(k)\leq\pi_{\bar{I}(k)}(k)$. From \eqref{es1}, we have
	\begin{equation}
		\begin{split}
			\pi_{\sigma(k)}(k)=&\underset{j\in\sigma(k)}{\max}\left| \hat{x}_{\sigma(k)}(k)-X_{ij}(k)\right| \\
			\geq&\left| \hat{x}_{\sigma(k)}(k)-X_{i\bar{j}(k)}(k)\right|.
		\end{split}
	\end{equation}
	Using the lower bound on $\pi_{\sigma(k)}(k)$ and the triangle inequality, we have that
	\begin{equation}\label{ee}
		\begin{split}
			|e_{\sigma(k)}(k)|=&|\hat{x}_{\sigma(k)}-x_{i}(k)|\\
			=&\left| \hat{x}_{\sigma(k)}(k)-X_{i\bar{j}(k)}(k)+X_{i\bar{j}(k)}(k)- x_{i}(k)\right| \\
			\leq&\pi_{\sigma(k)}(k)+|m_{\bar{j}}(k)|\\
			\leq&\pi_{\bar{I}(k)}(k)+|m_{\bar{j}}(k)|\\
			\leq&||m||_{\infty}+\underset{j\in \bar{I}(k)}{\max}|m_{j}(k)|+|m_{\bar{j}}(k)|\\
			\leq&3 ||m||_{\infty}.
		\end{split}
	\end{equation}
	Inequality \eqref{ee} is of the form \eqref{sa} and the result follows.\hfill$\blacksquare$\\[2mm]
	%\begin{assumption}\label{a2}
	%	For all $i\in I_{A}(k)$ and $k\geq 0$, among the set $\left\lbrace \mathcal{N}_{A}(k)\cap\mathcal{N}_{i}(k)\right\rbrace \cup\left\lbrace A,i\right\rbrace  $, the number of malicious vehicles does not exceed $\frac{N_{i}(k)}{2}$, i.e.,
	%	\begin{equation}
	%		\card(W_{i}(k))\leq q_{i}(k)<\frac{N_{i}(k)}{2},
	%	\end{equation}
	%	where $N_{i}(k)=2+\card(\mathcal{N}_{A}(k)\cap\mathcal{N}_{i}(k))$.
	%\end{assumption}
	If Assumptions \ref{as1}-\ref{as3} hold, for all $k\geq 0$ \eqref{es1}-\eqref{es3} can be repeated for every vehicle that belongs to $S(k)$ so that attack-free state estimation can be obtained. This estimation procedure is formally stated in Algorithm \ref{alg:the_alg1}.
	\begin{algorithm}[htb]
		\caption{Secure Estimation.}
		\label{alg:the_alg1}
		\begin{algorithmic}[1]
			\State For all $i\in S(k)$ and $k\geq 0$, collect $N_{i}(k)$ copies of $x_{i}(k)$ that are uploaded by all vehicles in $\mathcal{N}_{i+}(k)$.
			\State For all $i\in S(k)$ and $k\geq0$, perform \eqref{es1}-\eqref{es3} to obtain $\hat{x}_{i}(k)$.
			%\EndFor
			\State Return $\hat{x}_{i}(k)$ for all $i\in S(k)$ and $k\geq 0$.
			%\EndFor
		\end{algorithmic}
	\end{algorithm}

	\begin{remark}
		Because of the time-variance of $N_{i}(k)$, $q_{i}(k)$, $S(k)$ and $\mathcal{N}_{i}(k)$ for $i\in S(k)$, our estimation scheme allows vehicles to frequently enter and leave the cloud. 
	\end{remark}
	%\begin{remark}
	%	The advantage of Algorithm 2 over Algorithm 1 is that it confirms the physical existence of the vehicle before collecting the information it uploads so that Sybil attacks are less harmful. However, the time it takes for this confirmation might also bring delay to the estimation procedure.
	%\end{remark}
	%\begin{remark}
	%	If for some $i\in I_{A}(k)$ and $k\geq 0$, the number of the mutual surrounding vehicles of our ego vehicle and vehicle $i$ is large at time $k$, i.e., $\card(\mathcal{N}_{A}(k)\cap\mathcal{N}_{i}(k))$ is large, the dimension of $X(k)$, which is $N_{i}(k)$, will be large. This may largely increase the burder of computation for estimating $\hat{x}_{i}(k)$. In this case, we may randomly choose a subset $J_{max}\subset\left\lbrace 1,\ldots,N_{i}(k)\right\rbrace $ of information from $X(k)$ with $\card(J_{max})=N_{max}$, where the values of $N_{max}$ is determined by the computation power of our ego vehicle. Then we can perform \eqref{es1}-\eqref{es3} using the subset of information to obtain $\hat{x}_{i}(k)$ at time $k$.
	%\end{remark}
	\section{Malicious Vehicles Isolation}\label{isolation}
	In this section, we consider the design of an algorithm for isolating malicious vehicles.
	Suppose a priori knowledge of the noise bounds are known, we may perform isolation of malicious vehicles. For instance, we assume that the bound on the disturbance caused by network effects and sensor noise does not exceed $\gamma$, with value of $\gamma$ known, i.e.,
	\begin{equation}
		||m||_{\infty}=\gamma,
	\end{equation}
	and $\gamma$ is known.
	Under Assumptions \ref{as1}-\ref{as3}, we can obtain $\hat{x}_{i}(k)$ for all $i\in S(k)$ and $k\geq 0$ by running Algorithm 1. If vehicle $i$ is attack-free, then the information it uploads to the cloud, i.e., its own state and the state of its surrounding vehicles will be consistent with the corresponding attack-free estimates obtained by the cloud. Therefore, for all $i\in S(k)$ and $j\in\mathcal{N}_{i+}(k)$, $k\geq 0$, we compute the difference between $\hat{x}_{j}(k)$ and $\tilde{x}_{j|i}(k)$, where $\tilde{x}_{j|i}(k)=\tilde{x}_{i}(k)$ if $j=i$. From Theorem 2, we have proved that $|e_{i}(k)|
	\leq3||m||_{\infty}$ for all $i\in S(k)$. Then if vehicle $i$ is attack-free, for all $j\in\mathcal{N}_{i+}(k)$, we have
	%\begin{equation}\label{i1}
	%	\begin{split}
	%		|\hat{x}_{i}(k)-\tilde{x}_{i}(k)|=&|\hat{x}_{i}(k)-x_{i}(k)+x_{i}(k)-\tilde{x}_{i}(k)|\\
	%		\leq&|e_{i}(k)|+|m_{i}(k)|\\
	%		\leq&3||m(k)||_{\infty}+|m_{i}(k)|\\
	%		\leq&4\gamma;
	%	\end{split}
	%\end{equation}
	%similarly, the information of vehicle $j$ for all $j\in\mathcal{N}_{i}(k)\cap I_{A}(k)$ and $Ambiguity_{j}(k)=0$ that is provided by vehicle $i$, which is denoted as $x_{j|i}(k)$, will be consistent with $\hat{x}_{j}(k)$. We have
	\begin{equation}\label{i1}
		\begin{split}
			|\hat{x}_{j}(k)-\tilde{x}_{j|i}(k)|=&|\hat{x}_{j}(k)-x_{j}(k)+x_{j}(k)-x_{j|i}(k)|\\
			\leq&|e_{j}(k)|+|m_{i}(k)|\\
			\leq&3||m||_{\infty}+|m_{i}(k)|\\
			\leq&4\gamma.
		\end{split}
	\end{equation}
	Therefore, for every vehicle $i$ that belongs to $S(k)$, if \eqref{i1} holds for all $j\in\mathcal{N}_{i+}(k)$, vehicle $i$ is claimed to be attack-free at time $k$; otherwise, vehicle $i$ is malicious at time $k$. The above isolation procedure is summarized in Algorithm \ref{alg:the_alg2}.
	\begin{algorithm}[htb]
		\caption{Malicious Vehicle Isolation.}
		\label{alg:the_alg2}
		\begin{algorithmic}[1]
			\State Run Algorithm \ref{alg:the_alg1} to obtain $\hat{x}_{i}(k)$ for all $i\in S(k)$ and $k\geq 0$.
			%			\State \text{\textbf{If $\pi(k)>\bar{z}$, then}}
			%			\begin{equation*}
			%				detection(k)=1
			%			\end{equation*}
			%			\State \text{\textbf{else}}
			%			\begin{equation*}
			%			detection(k)=0
			%			\end{equation*}
			\State For all $i\in S(k)$ and $k\geq 0$, if \eqref{i1} holds for all $j\in\mathcal{N}_{i}(k)$,
			then vehicle $i$ is attack-free at time $k\geq 0$; otherwise, vehicle $i$ is malicious at time $k\geq 0$.
			\State The set of malicious vehicles in the cloud at time $k\geq 0$, which we denote as $\hat{W}_{A}(k)$ is given as
			\Statex
			$\hat{W}_{A}(k)=\lbrace i\in S(k)|\eqref{i1}$ is false for some $j\in\mathcal{N}_{i+}(k).$
			%				$\hat{W}_{A}(k)=\left\lbrace i\in S(k)|\eqref{i1}\hspace{2mm}is\hspace{2mm}false\hspace{2mm}for\hspace{2mm}some\hspace{2mm}j\in\mathcal{N}_{i+}(k)\hspace{2mm}with\hspace{2mm}Ambiguity_{j}(k)=0\right\rbrace .$
			%\EndIf
			%\EndFor
			\State Return $\hat{W}_{A}(k)$ for all $k\geq 0$.
			%\EndFor
		\end{algorithmic}
	\end{algorithm}	
	\section{Simulation}\label{simulation}
	In this section, numerical examples are provided to show the performance of our algorithms.
%			\begin{figure}
%		\includegraphics[width=0.5\textwidth]{x1estimate.eps}
%		\caption{Estimated $x$-coordinate $\hat{x}_{1}$ stays in a neighbourhood of $x_{1}$.}
%		\centering
%		\label{fig:0s}
%	\end{figure}
\begin{figure}
	\includegraphics[width=0.5\textwidth]{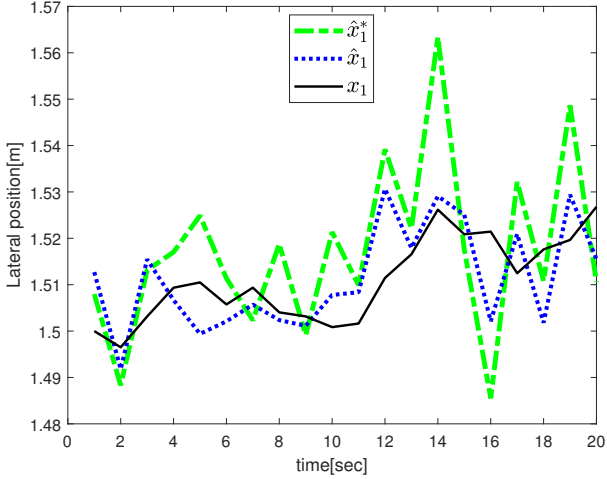}
	\caption{Estimated lateral position of vehicle 1 $\hat{x}_{1}$ given by Algorithm \ref{alg:the_alg1} and $\hat{x}_{1}^{*}$ given by \cite{golle2004detecting} stay in a neighbourhood of the true lateral position of vehicle $1$ $x_{1}$.}
	\centering
	\label{fig:6s}
\end{figure}
\begin{figure}
	\includegraphics[width=0.5\textwidth]{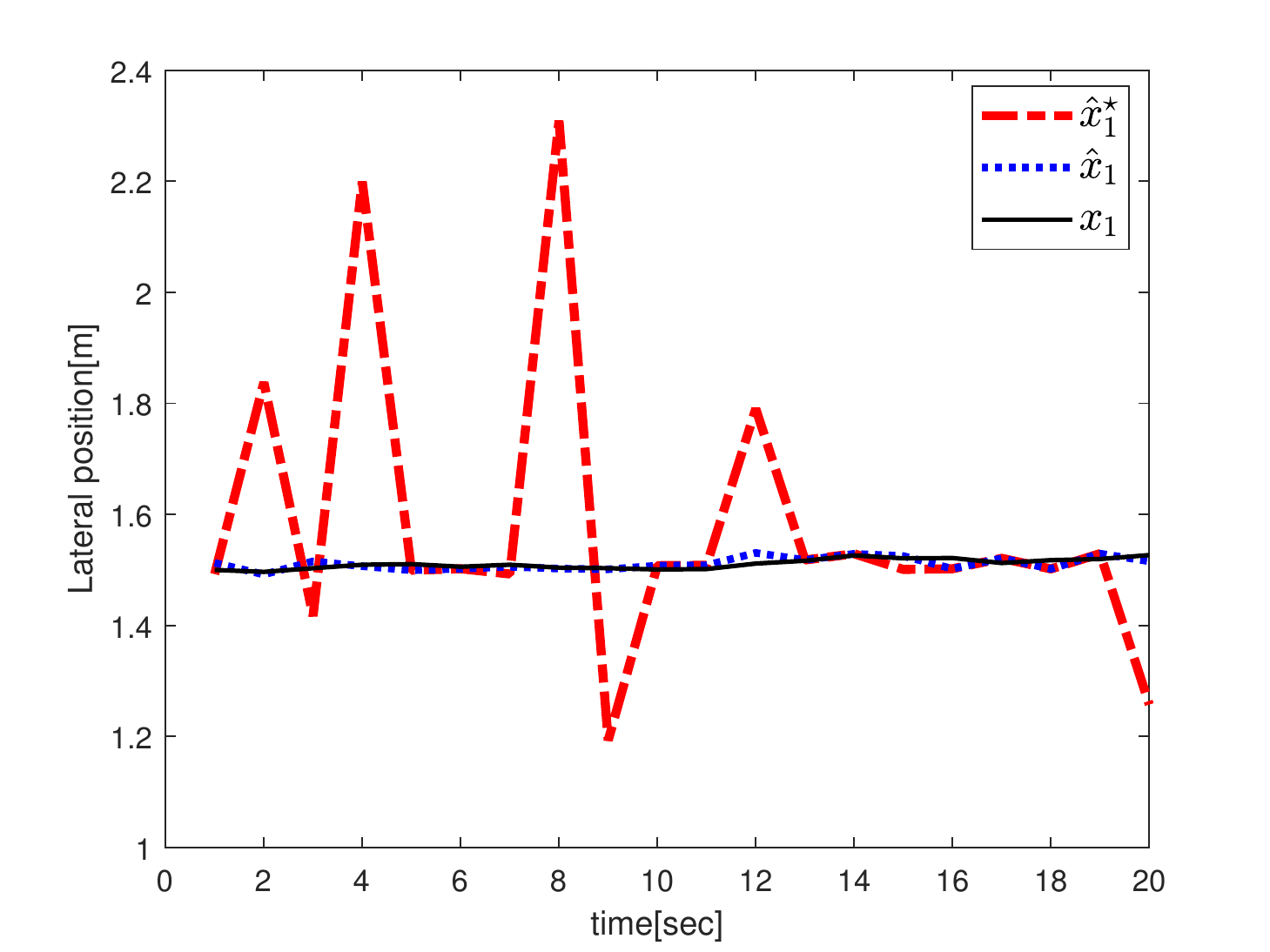}
		\caption{Estimated lateral position of vehicle 1 $\hat{x}_{1}$ given by Algorithm \ref{alg:the_alg1} and $\hat{x}_{1}^{\star}$ given by \cite{chattopadhyay2018secure} stay in a neighbourhood of the true lateral position of vehicle $1$ $x_{1}$.}
	\centering
	\label{fig:7s}
\end{figure}
%	\begin{figure*}[t]\centering
%		\includegraphics[width=0.9\textwidth]{simulationvehicle3.pdf}
%		\caption{$5$ vehicles and $3$ malicious identities in the cloud; vehicles 1-3 are attack-free (green); vehicles 4 and 5 are malicious (red); malicious identities 1-3 (grey); RSUs detect physical existence by generating radio signals to the location reported by each identity, i.e., if the identity exists physically, e.g., vehicles 1-5, echo returns from the reported location (solid arrow), otherwise, e.g., identitie 1-3, no echo returns from the reported location (dashed arrow).}
%		\centering
%		\label{fig:simulation}
%	\end{figure*}
		\begin{figure}
	\includegraphics[width=0.5\textwidth]{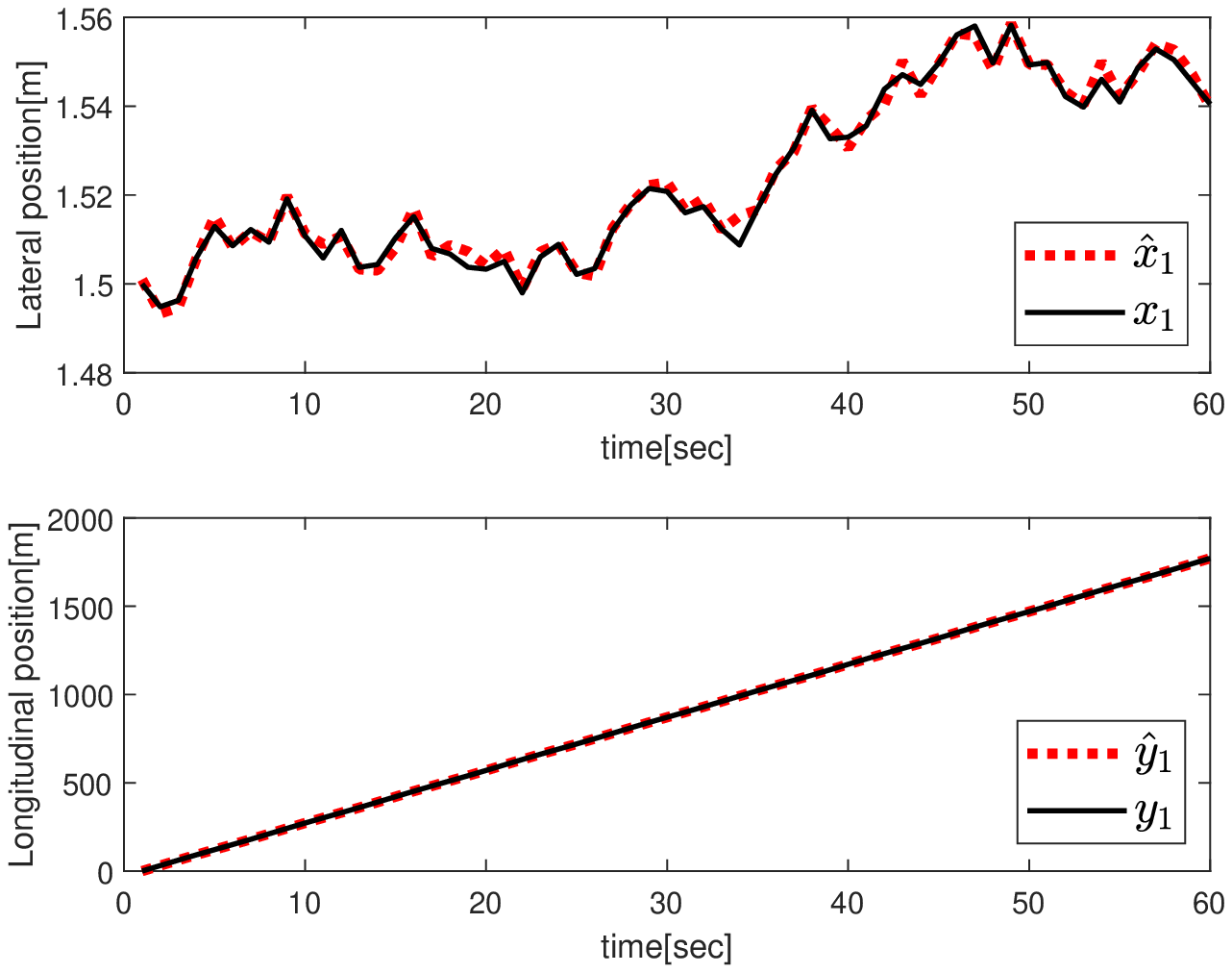}
	\caption{Estimated lateral and longitudinal positions of vehicle $1$ $\hat{x}_{1}$, $\hat{y}_{1}$ stays in the neighbourhood of the true lateral and longitudinal positions $x_{1}$, $y_{1}$ respectively.}
	\centering
	\label{fig:1s}
\end{figure}
\begin{figure}
	\includegraphics[width=0.5\textwidth]{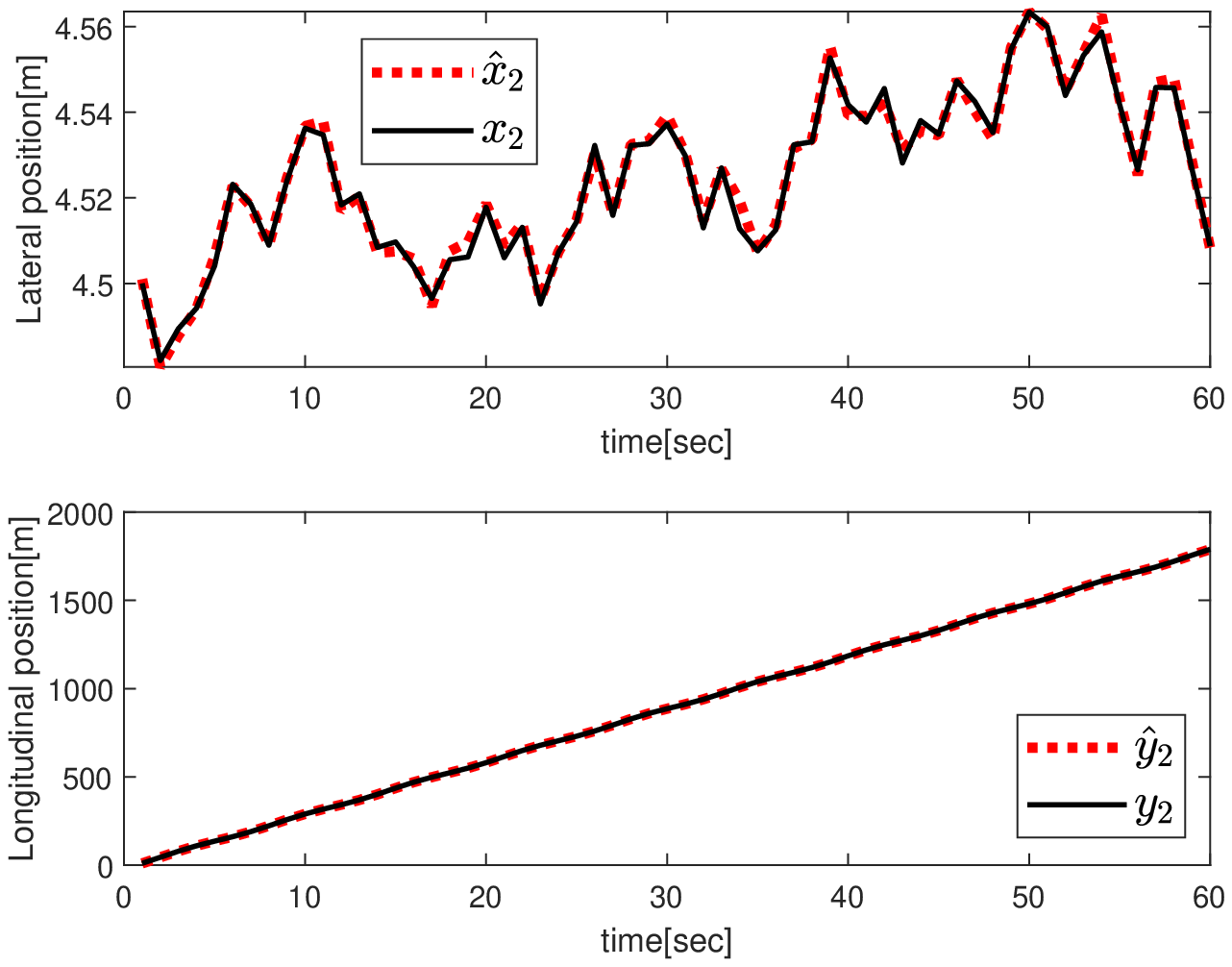}
	\caption{Estimated lateral and longitudinal positions of vehicle $2$ $\hat{x}_{2}$, $\hat{y}_{2}$ stays in the neighbourhood of the true lateral and longitudinal positions $x_{2}$, $y_{2}$ respectively.}
	\centering
	\label{fig:2s}
\end{figure}
\begin{figure}
	\includegraphics[width=0.5\textwidth]{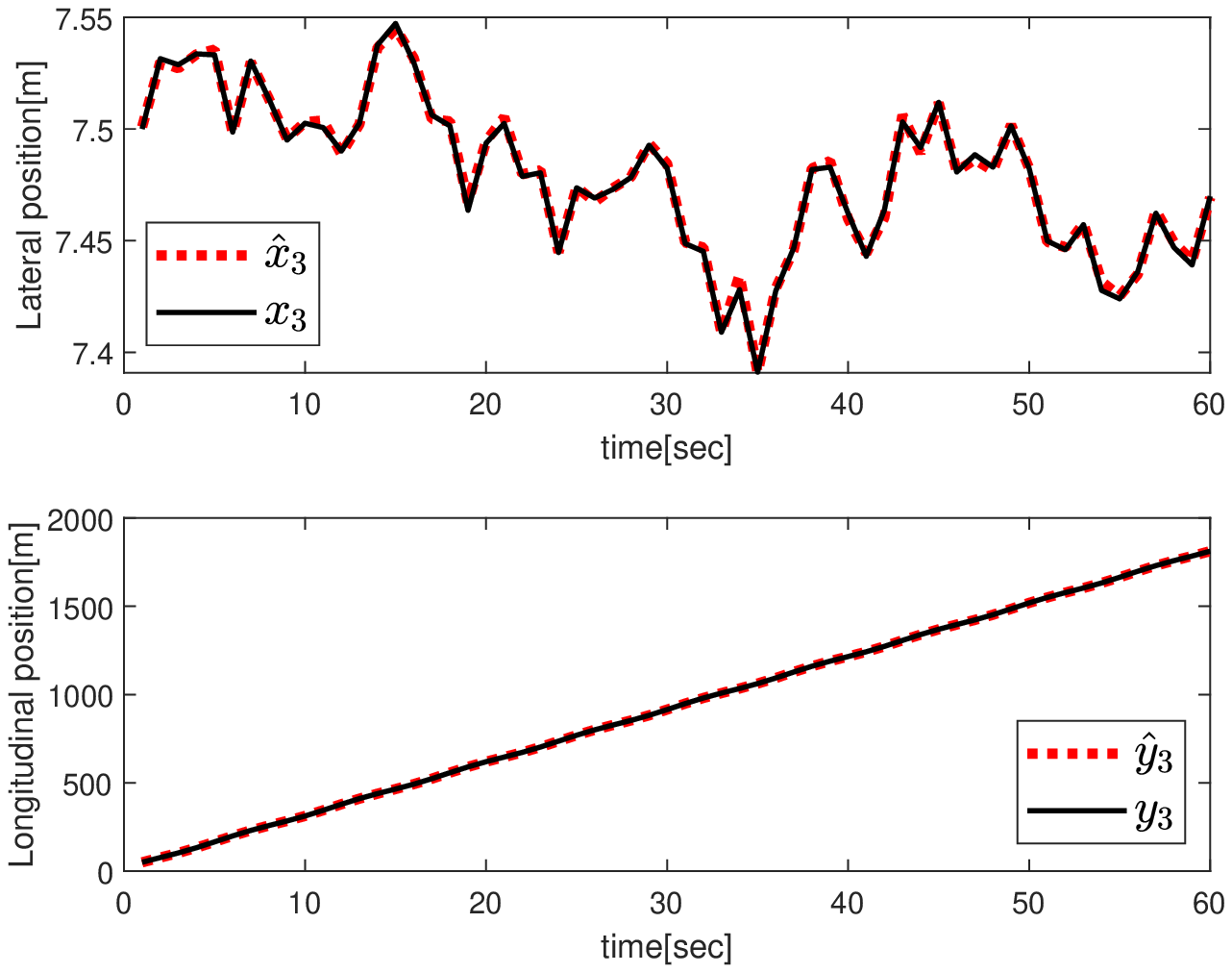}
	\caption{Estimated lateral and longitudinal positions of vehicle $3$ $\hat{x}_{3}$, $\hat{y}_{3}$ stays in the neighbourhood of the true lateral and longitudinal positions $x_{3}$, $y_{3}$ respectively.}
	\centering
	\label{fig:3s}
\end{figure}
\begin{figure}
	\includegraphics[width=0.5\textwidth]{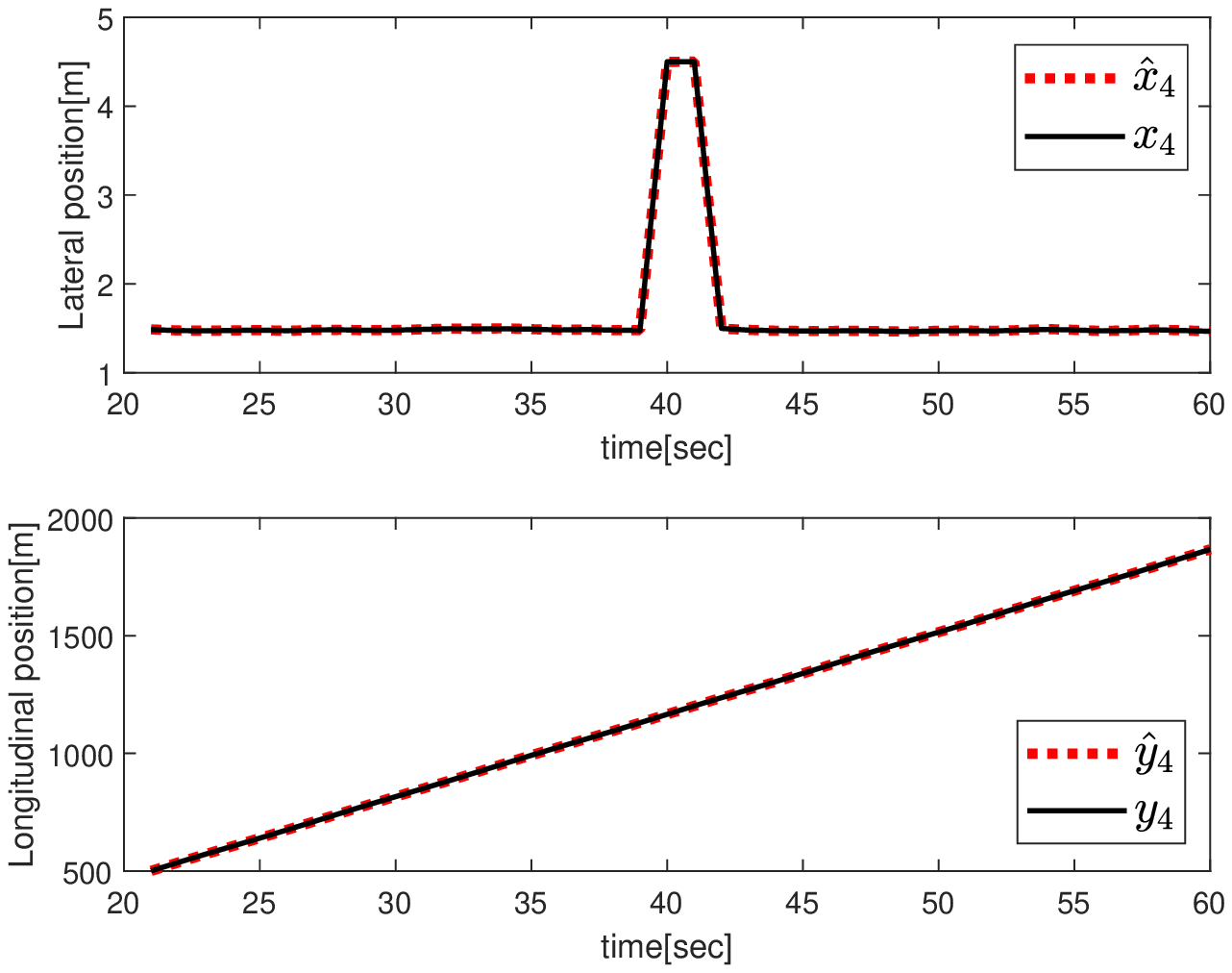}
	\caption{Estimated lateral and longitudinal positions of vehicle $4$ $\hat{x}_{4}$, $\hat{y}_{4}$ stays in the neighbourhood of the true lateral and longitudinal positions $x_{4}$, $y_{4}$ respectively.}
	\centering
	\label{fig:4s}
\end{figure}
\begin{figure}
	\includegraphics[width=0.5\textwidth]{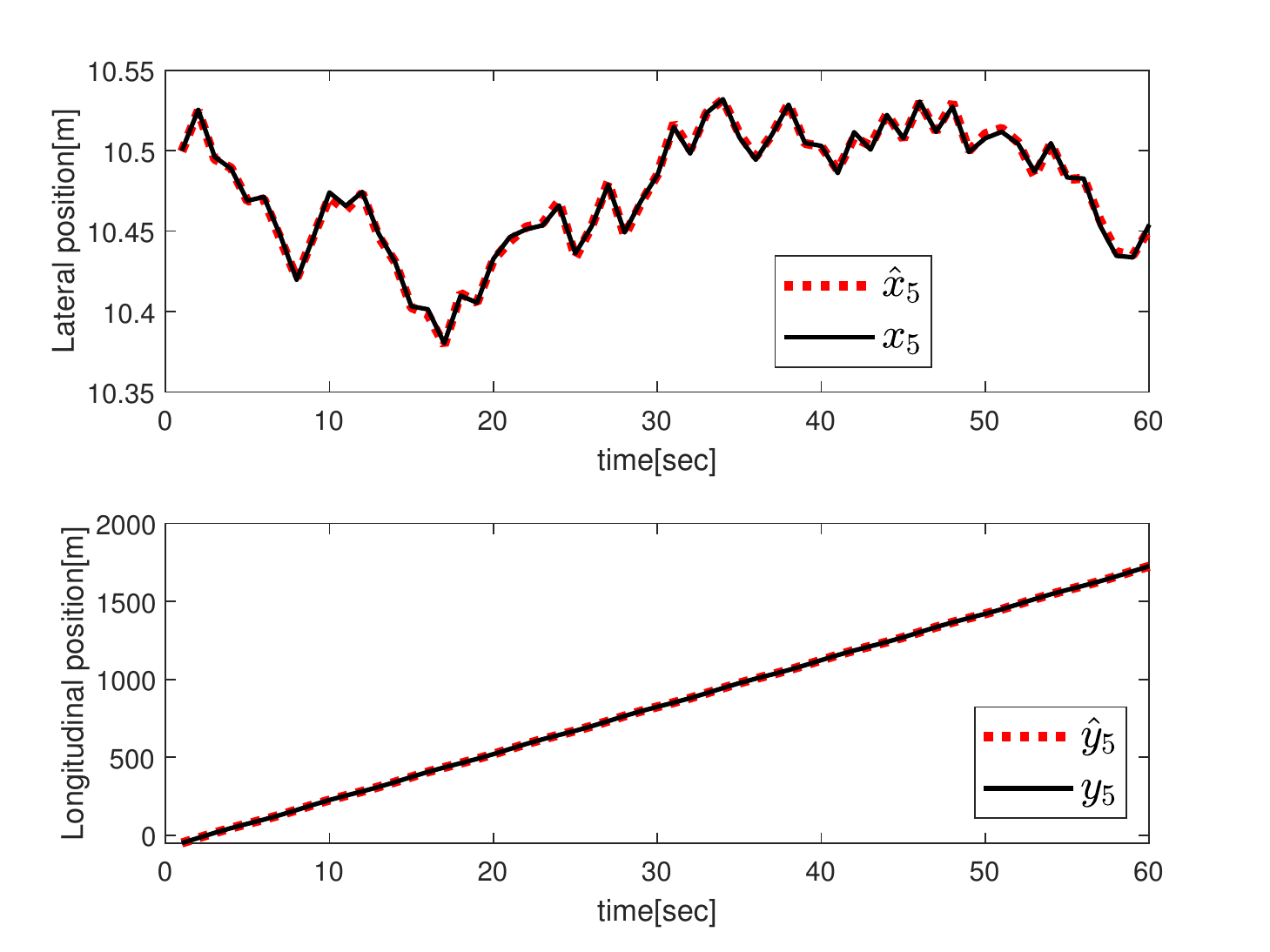}
	\caption{Estimated lateral and longitudinal positions of vehicle $5$ $\hat{x}_{5}$, $\hat{y}_{5}$ stays in the neighbourhood of the true lateral and longitudinal positions $x_{5}$, $y_{5}$ respectively.}
	\centering
	\label{fig:5s}
\end{figure}

\begin{figure}
	\includegraphics[width=0.5\textwidth]{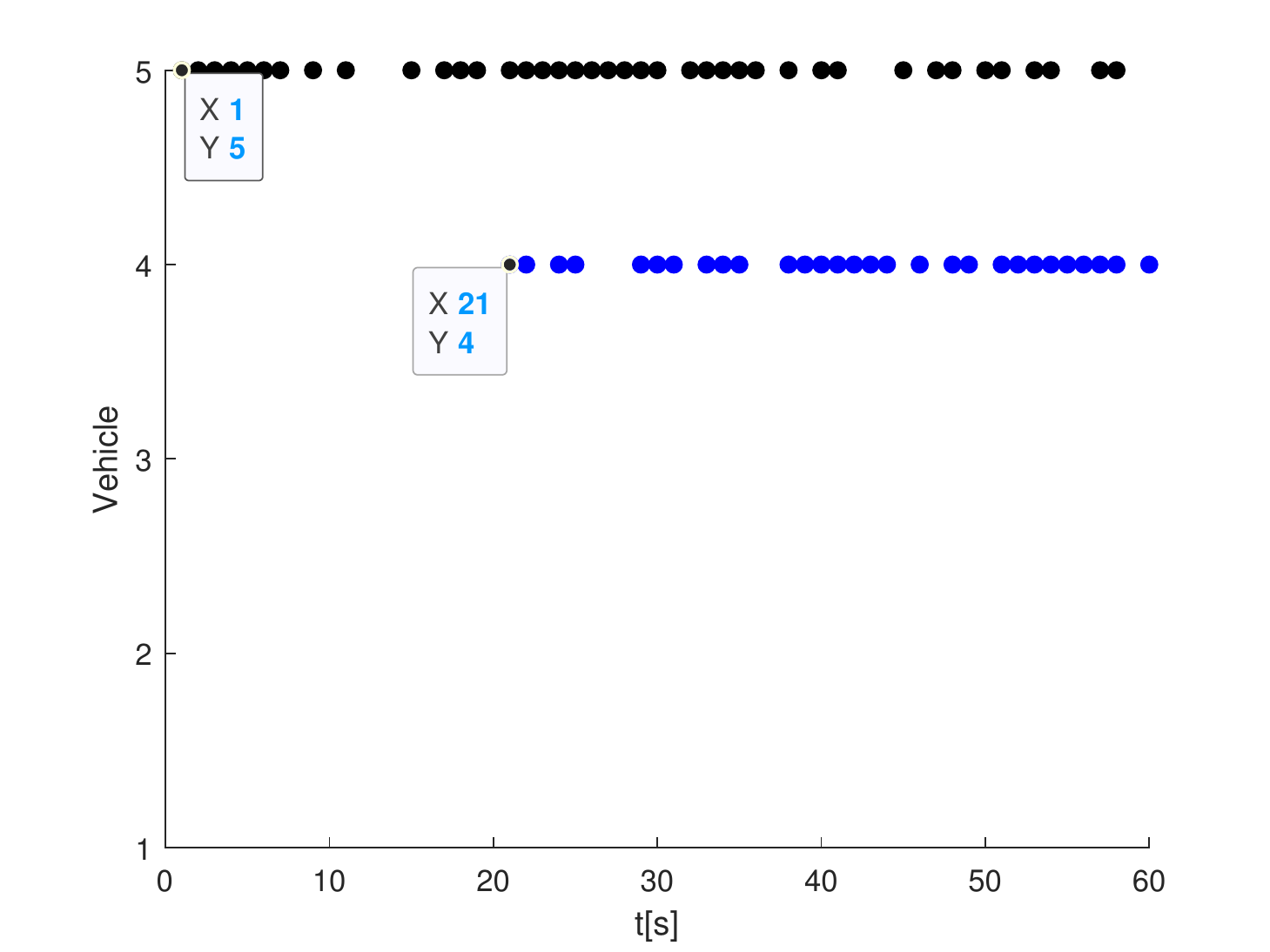}
	\caption{The isolated malicious vehicles, vehicle $4$ and $5$ start to be isolated when they belong to $S(k)$.}
	\centering
	\label{fig:11}
\end{figure}

	\textbf{Example 1}: We consider $5$ vehicles in the cloud, the lateral position of vehicle $1$ $x_{1}(t)=1.5+v_{1}(t)$ with $v_{1}(t)\in\mathcal{U}(-0.01,0.01)$ for all $t\geq 0$. At each time step $k$, vehicle $1$ uploads its own lateral position $x_{1}(k)$ to the cloud and vehicles $2,3,4,5$ are also measuring and uploading $x_{1}(k)$ to the cloud. Therefore, the cloud receives $5$ copies of $x_{1}(k)$ for $k\geq 0$. The measurement noise of each vehicle $m_{i}\in\mathcal{U}(-b_{i},b_{i})$ with $b_{i}=0.01*i$ for $i\in\left\lbrace 1,2,3,4,5\right\rbrace $. The cloud does not know the value of $b_{i}$ for any $i\in\left\lbrace 1,2,3,4,5\right\rbrace $. At each time step, suppose two of the vehicles are randomly selected to be malicious but we do not know which, and $a_{i}(k)\in\mathcal{N}(0,5^{2})$, for $i\in W_{1}(k)$. Assumptions \ref{as1}-\ref{as3} are satisfied, for $k\in[1,20]$, the cloud runs Algorithm \ref{alg:the_alg1} to estimate $x_{1}(k)$. The estimation given by Algorithm \ref{alg:the_alg1} is denoted as $\hat{x}_{1}$. For comparison, for $k\in[1,20]$, we also use the method given in \cite{golle2004detecting} for estimating $x_{1}(k)$, and the obtained estimation is denoted as $\hat{x}_{1}^{*}$. In particular, we adopt the idea of \cite{candes2005decoding} to find the best explanation of the received data by minimizing the difference between the estimated data and the received data. The performances of our approach and the one in \cite{golle2004detecting} is shown in Figure \ref{fig:6s}. We also compare our approach with the adaptive filtering approach in \cite{chattopadhyay2018secure} for estimating $x_{1}(k)$. Since $W_{1}(k)$ might be changing at every time step, the value of parameter $\tau$ in (7) of \cite{chattopadhyay2018secure} is set to be $1$, and the estimation given by \cite{chattopadhyay2018secure} is denoted as $\hat{x}_{1}^{\star}$. The performance is shown in Figure \ref{fig:7s}. 
%	\textbf{Example 1}: We consider $3$ vehicles in the cloud, the $x$-coordinate of vehicle $1$ $x_{1}(t)=1.5+v_{1}(t)$ with $v_{1}(t)\in\mathcal{U}(-0.01,0.01)$ for all $t\geq 0$. At each time step $k$, vehicle $1$ uploads its own $x$-coordinate $x_{1}(k)$ to the cloud and vehicles $2,3$ are also measuring and uploading $x_{1}(k)$ to the cloud. Therefore, the cloud receives $3$ copies of $x_{1}(k)$ for $k\geq 0$. The measurement noise of each vehicle $m_{i}\in\mathcal{U}(-b,b)$ with $b=0.005$ for $i\in\left\lbrace 1,2,3\right\rbrace $. The cloud does not know the value of $b$. At each time step, suppose one of the vehicles is randomly selected to be malicious but we do not know which, and $a_{i}(k)\in\mathcal{N}(0,5^{2})$ for some $i\in\left\lbrace 1,2,3\right\rbrace $. Assumptions \ref{as1}-\ref{as3} are satisfied, for $k\in[1,20]$, the cloud runs Algorithm \ref{alg:the_alg1} to estimate $x_{1}(k)$ and the performance is shown in Figure \ref{fig:0s}. 

	\textbf{Example 2}: We consider $5$ vehicles in the cloud, running on $4$ lanes of a highway in the same direction. Vehicles $4$ and $5$ are malicious. Our cloud does not know how many or which vehicles in the cloud are malicious. The initial positions of the $5$ vehicles are $(1.5,0)$, $(4.5,10)$, $(7.5,50)$, $(1.5,-200)$, $(10.5,-50)$. Their longitudinal speeds with unit $(m/s)$ are given as $v_{y1}(t)=30+\sin(t)$, $v_{y2}(t)=30+5\sin(t)$, $v_{y3}(t)=30-4\sin(t)$, $v_{y4}(t)=35+\sin(t)$, $v_{y5}(t)=30+3\sin(t)$; and lateral speeds $v_{x1}(t)\in\mathcal{U}(-0.01,0.01)$, $v_{x2}(t)\in\mathcal{U}(-0.02,0.02)$, $v_{x3}(t)\in\mathcal{U}(-0.04,0.04)$, $v_{x4}(t)\in\mathcal{U}(-0.01,0.01)$, $v_{x5}(t)\in\mathcal{U}(-0.03,0.03)$ when vehicles are not changing their lanes. The noise in the lateral position measurement of each vehicle $m_{xi}(k)\in\mathcal{U}(-0.005,0.005)$ for all $i\in\left\lbrace 1,2,3,4,5\right\rbrace $; the noise in the longitudinal position measurement of each vehiccle $m_{yi}(k)\in\mathcal{U}(-0.5,0.5)$ for all $i\in\left\lbrace 1,2,3,4,5\right\rbrace $. Therefore, $\gamma=0.005$ for lateral position measurement, and $\gamma=0.5$ for longitudinal position measurement. For all $k\geq 0$, each vehicle and malicious identity upload its own position and its relative position with the other vehicles to the cloud. Let the time interval between every two consecutive time steps be $1$ second. The attack signals injected in vehicles $4$ and $5$ for all $k\geq 0$ are given as $a_{4}(k)\in\mathcal{N}(0,5^{2})$ and $a_{5}(k)\in\mathcal{N}(0,5^{2})$, which are added to all the information they upload to the cloud. 
	
	We assume for $k\in [1,60]$, $S(k)$ is the set of vehicles within $100$ meters range from vehicle $1$. 
	
	\textit{Period 1}: For $k\in [1,20]$, vehicles $1$, $2$, $3$, $5$ stay in this range and vehicle $4$ is out of ths range. Hence, $S(k)=\left\lbrace 1,2,3,5\right\rbrace $ in this period. Since the speed of vehicle $4$ is larger than vehicle $1$, the distance between them becomes smaller as time grows.
	
	\textit{Period 2}: For $k\in[21,60]$, vehicle $4$ enters the range at $k=21$ and it overpasses vehicle $1$ from $k=39$ to $k=42$ so there are lane changes during that period. Vehicle $4$ leaves the range when $k>60$. Hence, $S(k)=\left\lbrace 1,2,3,4,5\right\rbrace $ in period 2.
	
	For all $i\in S(k)$ and $k\in [1,60]$, Assumptions \ref{as1}-\ref{as3} are satisfied, without knowing the values of $\gamma$, Algorithm \ref{alg:the_alg1} is performed for estimating the lateral and longitudinal positions of vehicle $i$ for $i\in S(k)$, which we denote as $x_{i}(k)$ and $y_{i}(k)$ respectively for $i\in S(k)$. The corresponding estimates are denoted as $\hat{x}_{i}(k)$ and $\hat{y}_{i}(k)$ respectively for $i\in S(k)$. We depict the performance of Algorithm \ref{alg:the_alg1} in Figures \ref{fig:1s}-\ref{fig:5s}.
	
	If the values of $\gamma$ are known for the lateral and longitudinal position measurement noise, then Algorithm \ref{alg:the_alg2} can be performed for isolating malicious vehicles in $S(k)$. The isolation result is shown in Figure \ref{fig:11}, where vehicles $4$ and $5$ are successfully isolated as malicious ones. Note that vehicle $5$ belongs to $S(k)$ for all $k\in[1,60]$, while vehicle $4$ starts to be isolated as malicious after $20$ seconds since it belongs to $S(k)$ for $k\in [21,60]$. It has been shown that vehicle $5$ becomes isolated at time $1$, and vehicle $4$ becomes isolated at time $21$, i.e., the malicious vehicles are isolated as soon as it becomes an element of set $S(k)$, which shows the efficiency of Algorithm \ref{alg:the_alg2}.

	\section{Conclusion}\label{conclusion}
	Exploiting the information redundancy on the cloud side created by each vehicle sharing its local sensor measurements, we have proposed a sensor fusion algorithm that provides robust state estimation of all vehicles in the cloud when the number of malicious vehicles is sufficiently small and sensor disturbances are present. Unlike other relevant works, we do not assume any a prior knowledge of sensor disturbance is available except its boundedness when performing state estimation. We use these estimates for isolating malicious vehicles in the cloud.

	\bibliographystyle{ieeetr}
	\bibliography{Observer1}
\end{document}